\documentclass[aip,reprint,pop]{revtex4-1}

\usepackage{graphicx}
\usepackage{subfigure}
\usepackage{epstopdf}
\usepackage{amsmath}
\usepackage{amssymb}
\usepackage{nicefrac}
\usepackage{dcolumn}
\usepackage{bm}
\usepackage[colorlinks=true, linkcolor=blue]{hyperref}

\begin{document}

\title{Geometric phase in Brillouin flows}
\author{Jean-Marcel Rax}
\affiliation{D\'epartement de Physique, Universit\'{e} de Paris XI - Ecole Polytechnique, 91128 Palaiseau, France}
\author{Renaud Gueroult}
\affiliation{LAPLACE, Universit\'{e} de Toulouse, CNRS, INPT, UPS, 31062 Toulouse, France}

\begin{abstract}
A geometric phase is found to arise from the cyclic adiabatic
variation of the crossed magnetic and electric fields which sustain the Brillouin rotation of a plasma
column. The expression of the gauge field associated with this geometric phase
accumulation is explicited. The physical origin of this phase is shown to be the uncompensated inductive electric field drift that stems from magnetic field cyclic variations. Building on this result, the effect of a weak,
periodic and adiabatic modulation of the axial magnetic field on the particle guiding center drift motion is demonstrated to be equivalent to that of a perpendicular electric field, allowing to study the gauge induced Brillouin flow through a geometrically equivalent linear radial electric field. This finding opens new perspectives to drive plasma rotation and hints at possible applications of this basic effect.
\end{abstract}

\date{\today}
\maketitle

\section{Introduction}

The Brillouin rotation~\cite{Brillouin1945,Gueroult2013} of cylindrical beams and plasma columns produced by the combination of a uniform axial magnetic field and a linear radial electric field has been
previously investigated in plasma centrifuges~\cite{Bonnevier1966,Krishnan1981,Prasad1987,Ohkawa2002,Shinohara2007,Fetterman2011,Gueroult2014,Gueroult2019} and magnetic confinement fusion experiments including tokamaks~\cite{Rax2017} and rotating mirrors~\cite{Wilcox1959,Lehnert1971,Bekhtenev1980,Fetterman2008}, as well as in basic beam and nonneutral plasmas physics studies~\cite{Davidson2001}. The complementary problem of
angular momentum conversion to produce rotation has also been largely
studied, and the general problem of angular momentum conversion between \
electromagnetic fields and particles has for instance received considerable
attention in the study of (\textit{i}) dissipation in tokamak and
mirrors~\cite{Fetterman2011a,Rax2017,Rax2019}, (\textit{ii}) particle acceleration~\cite{Rax2010,Thaury2013}, (\textit{%
iii}) magnetic field generation~\cite{Shvets2002,Kostyukov2002} and (\textit{iv}) isotopes
separation~\cite{Dolgolenko2009,Rax2007,Gueroult2014a,Rax2015,Rax2016}. Yet, and despite this broad range of fundamental and applied studies of the dynamics of magnetized plasma
rotation, the possible emergence of a geometric phase in crossed
field rotating Brillouin flows, as well as the question of the application
of this geometric phase, if any, have received far less interest.

Consider an integrable Hamiltonian system with at least two control
parameters. When these control parameters are constant the dynamics is
described by the invariance of the action variables and the rotation of the
angles variables~\cite{Goldstein1980}. When the two control parameters are slowly varying
along a closed curve $C$ in the control parameter space, \textit{phase effects} can be observed. During the \textit{cyclic adiabatic evolution} of the two Hamiltonian
control parameters along $C$, the action variables remain unchanged, but
the angles displays two types of phase accumulation. First, a \textit{%
dynamical} phase which increases with the \textit{duration} of the cyclic
evolution. Second, a \textit{geometric} phase which increases with the 
\textit{surface} of the closed circuit $C$ in the control parameter space~\cite{Berry1984,Hannay1985,Berry1988}. Geometric phase effects are well documented in charged particle physics~\cite{Littlejohn1988,Zhu2017}. They have, for instance, been studied for the cyclotron motion~\cite{Liu2011,Brizard2012}, wave propagation and Faraday effects~\cite{Liu2012} and the bounce motion of trapped
particles~\cite{Burby2013}. 

Here, we identify and analyze a new geometric phase in a classical plasma
configuration, the Brillouin flow. A Brillouin rotation is controlled by two vectorial parameters: an axial magnetic field $\mathbf{B}\in\mathbb{R}^{3}$ and a radial electric fields $\mathbf{E}\in \mathbb{R}^{3}$. One can then expect a geometric phase effect in the presence of a slow cyclic variation of these parameters along a closed curve in $\mathbb{R}^{6}$. This is indeed true if the direction of
these fields changes along an adiabatic closed path in $\mathbb{R}^{6}$. This
type of geometric phase is called an Hannay angle. In contrast, we
explore in this study the effect of an adiabatic change of the
amplitude of these fields. The control parameters manifold is
extended to its fiber bundle, the control parameter phase space, to
identify a new geometric phase associated with an adiabatic closed path in 
$\mathbb{R}^{2}$. Beyond its fundamental interest, the choice to consider a variation of the fields
amplitude rather than of their directions is motivated by the fact that it is far simpler to
implement experimentally. This should in turn facilitate the observation of these effects as well as their applications.

This study is organized as follows. In Sec.~\ref{Sec:SecII} we extend the
classical analysis of Brillouin rotations to the adiabatic regime when the
electric and magnetic fields are slowly varying. This solution of the
adiabatic Brillouin problem is carried up to the third order in the
adiabaticity parameter. The solution obtained in Sec.~\ref{Sec:SecII} for the particle
orbits is then used in Sec.~\ref{Sec:SecIII} to identify the geometric phase associated
with the drift motion of its guiding center around the magnetic axis. We then
derive the expression of the gauge field which quantifies the accumulation of a
geometric phase. This phase is shown to increase with the surface of the closed circuit
in phase space which is travelled in the course of the cyclic adiabatic evolution of the radial electric field
and axial magnetic field. This new result is then analyzed and interpreted in Sec.~\ref{Sec:SecIV} in light of of electric drifts theory. For a cyclic adiabatic variation of the magnetic
field, the geometric phase rotation of the guiding center is found to be
due to a an uncompensated series of $\mathbf{E}_i$ cross $\mathbf{B}$ drifts and polarization
drifts where $\mathbf{E}_i$ is induced by the slow variations of $\mathbf{B}$.
This analysis of the geometric phase is then used in Sec.~\ref{Sec:SecV} \
to uncover a mechanism to drive plasma rotation without the need for a DC
electric field. The steady state adiabatic cyclic
evolution of the axial magnetic field drives an average guiding center rotation
which, in contrast with the Brillouin slow rotation mode, is dispersive with respect to
the particle mass for a given kinetic energy. Finally, the main results of this study are summarized in Sec.~\ref{Sec:SecVI}.

\section{Adiabatic theory of Brillouin rotation}
\label{Sec:SecII}

In this section we extend the classical static Brillouin solutions to the
adiabatic slowly varying regime up to third order in adiabaticity
parameter. In this study $\left( \mathbf{u}_{x},\mathbf{u}_{y},\mathbf{u}_{z}\right)$ designates a Cartesian basis and $\left( x,y,z\right)$ the associated Cartesian coordinates.

\begin{figure}
\begin{center}
\includegraphics[width=7cm]{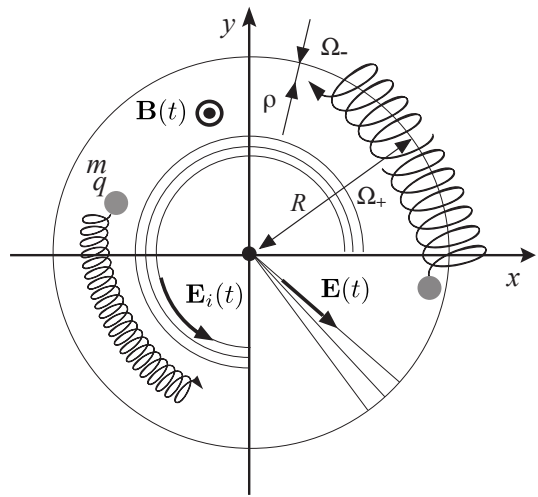}
\caption{ Basic field configuration for the adiabatic Brillouin problem: $%
\mathbf{E}$ is the capacitive radial electric field, $\mathbf{E}_{i}$ the
inductive azimuthal electric field and $\mathbf{B}$ the axial magnetic
field. Typical orbits in static fields illustrate the combination of
the fast cyclotron rotation $\left( \Omega _{-},\rho \right) $ and the slow
drift rotation $\left( \Omega _{+},R\right) $.}
\label{Fig:Fig1}
\end{center}
\end{figure}

Consider as illustrated in Fig.~\ref{Fig:Fig1} a charged particle of mass $m$ and charge $q$ in the electromagnetic field configuration made of :
\begin{subequations}
\begin{itemize}
\item[(\textit{i})] a time dependent homogeneous axial magnetic field 
\begin{equation}
\mathbf{B}\left( t\right) =\frac{m}{q}{\omega_{B}}\left( t\right) \mathbf{u}%
_{z}\text{,}  \label{1}
\end{equation}
which provides radial confinement, 
\item[(\textit{ii})] a radial capacitive electric field 
\begin{equation}
\mathbf{E}\left( \mathbf{r},t\right) =-\frac{m}{q}{\omega_{E}}^{2}\left(
t\right) \left( x\text{ }\mathbf{u}_{x}+y\text{ }\mathbf{u}_{y}\right) \text{%
,}  \label{2}
\end{equation}
which provides the $\mathbf{E}$ cross $\mathbf{B}$ drive for the Brillouin rigid body
rotation, 
\item[(\textit{iii})] an azimuthal inductive electric field 
\begin{equation}
\mathbf{E}_{i}\left( \mathbf{r},t\right) =\frac{1}{2}\frac{m}{q}\frac{%
d{\omega_{B}}}{dt}\left( y\text{ }\mathbf{u}_{x}-x\text{ }\mathbf{u}%
_{y}\right) \text{.}  \label{3}
\end{equation}
induced by the time variation of the magnetic field $\mathbf{B}$ in Eq.~(\ref{1}).
\end{itemize}
\label{123}
\end{subequations}
Here we assume that both ${\omega_{B}}\left( t\right) >0$ and $\omega_{E}\left( t\right) >0$, and $\mathbf{E}_{i}$ is written as a function of $\mathbf{B}$ through Maxwell-Faraday relation $\bm{\nabla }\times \mathbf{E}_{i}=-\partial \mathbf{B/}\partial t$. Although the ability to produce the radial electric field given in Eq.~(\ref{2}) remains an open question~\cite{Gueroult2019}, it can in principle be done in two ways. One conceptual solution is to use a set of segmented concentric electrodes acting as DC polarized end plate intercepting magnetic field lines~\cite{Lehnert1970}. Alternatively, the resonant absorption of an RF wave on a minority population can provide a steady state orbital angular momentum input to compensate angular momentum dissipation. In both cases the RF or DC power is used to counteract the relaxation of a small homogeneous space charge unbalance which is the source of $\mathbf{E}$.

In the field configuration given in Eqs.~(\ref{1}), (\ref{2}) and (\ref{3}), the motion along magnetic field lines is a simple uniform translation, while the dynamics around and across magnetic field lines is given by the solution of the two equations of motion under the electric and magnetic
forces,
\begin{subequations}
\begin{align}
\frac{d^{2}x}{dt^{2}} &= {\omega_{B}}\left( t\right) \frac{dy}{dt}-{\omega_{E}}^{2}\left( t\right) x+\frac{1}{2}\frac{d{\omega_{B}}}{dt}y\text{,}
\label{4a} \\
\frac{d^{2}y}{dt^{2}} &=-{\omega_{B}}\left( t\right) \frac{dx}{dt}-{\omega_{E}}^{2}\left( t\right) y-\frac{1}{2}\frac{d{\omega_{B}}}{dt}x\text{.}
\label{4b}
\end{align}
\end{subequations}
The Brillouin solutions describe the case where fields are time independent, \emph{i.~e.} ${\omega_{B}}^{\prime }={{\omega_{E}}}^{\prime }=0$. In this case the orbits are trochoidal orbits as illustrated in Fig.~\ref{Fig:Fig1}. These orbits are a combination of (\textit{i}) a fast rotation associated with the cyclotron
motion (radial excursion $\rho$ in Fig.~\ref{Fig:Fig1}), and (\textit{ii}) a slow rotation
associated with the $\mathbf{E}$ cross $\mathbf{B}$ drift (orbit of radius $R$ in Fig.~\ref{Fig:Fig1}). 

The linear system Eqs.~(\ref{4a}-\ref{4b}) can be rewritten as a single linear equation
\begin{equation}
\frac{d^{2}Z}{dt^{2}}+\Omega ^{2}\left( t\right) Z=0\text{,}  \label{6}
\end{equation}
for the complex variable $Z\left( t\right)$ defined by 
\begin{equation}
\left[ x(t)+jy(t)\right] =Z(t)\exp \left[-j\int_{0}^{t}\frac{\omega_{B}(u)}{2}du\right]
\label{5}
\end{equation}
and the positive real frequency
\begin{equation}
\Omega \left( t\right) =\sqrt{{\omega_{E}}^{2}\left( t\right) +{\omega_{B}}^{2}\left( t\right) /4}\text{.}  \label{7}
\end{equation}
Using one additional variable transform
\begin{equation}
Z\left( t\right) =Z_{0}\exp \int_{0}^{t}\varphi \left( u\right) du\text{,}
\label{A2}
\end{equation}
we can solve Eq.~(\ref{6}) for the complex phase 
\begin{equation}
\varphi \left( t\right) = \alpha \left( t\right)  + j\beta \left( t\right)
\end{equation}
with $(\alpha,\beta)\in\mathbb{R}^{2}$. With this notation, $\alpha$ and $\beta$ describes the evolution of the amplitude and the phase of $Z$, respectively, and $Z_{0}$ and $\varphi \left( 0\right) Z_{0}$ are given by the initial values $Z\left( 0\right) $ and $Z^{\prime }\left( 0\right) $. The unknown
complex function $\varphi \left(t\right) $ then verifies a \textit{Ricatti
equation}, 
\begin{equation}
\frac{d\varphi }{dt}+\varphi ^{2}+\Omega ^{2}=0\text{,}  \label{A3}
\end{equation}
with the initial value $\varphi \left( 0\right)=Z^{\prime }\left( 0\right) /Z\left( 0\right)$.

We now assume that ${\omega_{B}}$ and ${\omega_{E}}$ are slowly varying and that the adiabaticity parameter $\varepsilon $ verifies the condition 
\begin{equation}
\varepsilon =\left| \Omega ^{\prime }\right| /\Omega ^{2}\ll 1\text{,}
\label{8}
\end{equation}
The small adiabaticity parameter $\varepsilon$ is then used to write the solution of Eq.~(\ref{A3}) as a series expansion in power of $\varepsilon$,
\begin{equation}
\varphi  = \varphi _{0} + \varphi _{1} + \varphi _{2} + \varphi _{3} + O\left[ \varepsilon ^{4}\right].
\end{equation}
Substituting this expansion in Eq.~(\ref{A3}) with $d/dt=O\left[ \varepsilon \right]$ and
ordering the terms in powers of $\varepsilon $ yields the classical recursive relation for both the asymptotic adiabatic theory of particles dynamics and the WKB theory of wave dynamics,
\begin{subequations}
\begin{align}
\varphi _{0}^{2}+\Omega ^{2} =&0\text{,}  \label{A8} \\
\frac{d\varphi _{0}}{dt}+2\varphi _{1}\varphi _{0} =&0\text{,}  \label{A145}
\\
\frac{d\varphi _{1}}{dt}+\varphi _{1}^{2}+2\varphi _{0}\varphi _{2} =&0%
\text{,}  \label{A9} \\
\frac{d\varphi _{2}}{dt}+2\varphi _{1}\varphi _{2}+2\varphi _{0}\varphi _{3}
=&0\text{.}  \label{A10}
\end{align}
\end{subequations}
The first three equations, Eqs.~(\ref{A8}), (\ref{A145}) and (\ref{A9}), are solved to write the first three terms of the adiabatic asymptotic expansion,
\begin{subequations}
\begin{align}
\varphi _{0} =&\pm j\Omega \text{, }\\
\varphi _{1}=&-\frac{\Omega ^{\prime }}{2\Omega },  \label{A13} \\
\varphi _{2} =&\mp j\frac{\Omega ^{\prime \prime }}{4\Omega ^{2}}\pm j\frac{%
3\Omega ^{\prime 2}}{8\Omega ^{3}}\text{,}  \label{A14}
\end{align}
\end{subequations}
and expressions can be derived similarly for higher order terms. The structure of the
recursive relation, Eqs.~(\ref{A13}) and (\ref{A14}), is such that all the odd
terms with respect to the exponent of $\varepsilon $ are real, while all the even terms with
respect to the exponent of $\varepsilon $ are imaginary. One can thus separate odd and even
terms as the real and imaginary parts of $\varphi $ = $\alpha $ + $j\beta $
where $\alpha $ is the sum of the odd terms and $j\beta $ the sum of the
even terms. Plugging $\varphi $ = $\alpha $ + $j\beta $ into
the Ricatti equation Eq.~(\ref{A3}), one gets
\begin{equation}
\frac{d\alpha }{dt}+j\frac{d\beta }{dt}+\left( \alpha +j\beta \right)
^{2}+\Omega ^{2}=0\text{.}  \label{A15}
\end{equation}
The imaginary part of Eq.~(\ref{A15}) can then be integrated to provide an additional
relation between $\alpha $ and $\beta $, 
\begin{equation}
\int_{t_{0}}^{t}\alpha \left( u\right) du=\ln \sqrt{\frac{\left| \beta
\left( t_{0}\right) \right| }{\left| \beta \left( t\right) \right| }}\text{.}
\label{A16}
\end{equation}
This relation Eq.~(\ref{A16}) is true to all $\varepsilon $ order and is
used here to express the odd terms $\left( \alpha \right) $ as a function of the
even ones $\left( \beta \right)$. The two independent
adiabatic solutions $Z_{\pm }$ of the original linear equation Eq.~(\ref{6}) finally write
\begin{eqnarray}
\frac{Z_{\pm }\left( t\right) }{Z_{\pm }\left( 0\right) } &=&\frac{\sqrt{%
\left| \Omega +\frac{3\Omega ^{\prime 2}}{8\Omega ^{3}}-\frac{\Omega
^{\prime \prime }}{4\Omega ^{2}}\right| _{0}}}{\sqrt{\left| \Omega +\frac{%
3\Omega ^{\prime 2}}{8\Omega ^{3}}-\frac{\Omega ^{\prime \prime }}{4\Omega
^{2}}\right| _{t}}}  \label{A19} \\
&&\times \exp \pm j\int_{0}^{t}\left( \Omega +\frac{3\Omega ^{\prime 2}}{%
8\Omega ^{3}}-\frac{\Omega ^{\prime \prime }}{4\Omega ^{2}}\right) du\text{,}
\nonumber
\end{eqnarray}
These solutions are accurate to third order in $\varepsilon$, that is to say beyond the classical first order adiabatic (WKB) expansion. Note also that there is no need here to explore higher order terms here since it is sufficient to identify the geometric phase in the tangent bundle $%
\left( \Omega ,\Omega ^{\prime }\right) $ of the control parameter space.

Under the adiabatic condition given in Eq.~(\ref{8}), the third order adiabatic
solution writes
\begin{multline}
Z\left( t\right) =\sum_{\pm }\frac{A_{\pm }}{\sqrt{\left| \Omega +\frac{%
3\Omega ^{\prime 2}}{8\Omega ^{3}}-\frac{\Omega ^{\prime \prime }}{4\Omega
^{2}}\right| _{t}}} \\
\times \exp \pm j\int_{0}^{t}\left( \Omega +\frac{3\Omega ^{\prime 2}}{%
8\Omega ^{3}}-\frac{\Omega ^{\prime \prime }}{4\Omega ^{2}}\right) du\text{.}
\end{multline}
It is a linear combination of the $Z_{\pm }$ functions given in Eq.~(\ref{A19}), with the coefficients $A_{\pm }$ determined by the initial conditions. Extending the classical definitions of the Brillouin slow $\left( \Omega
_{+}\right) $ and fast $\left( \Omega _{-}\right) $ angular frequencies from the
static to the adiabatic regime, we define
\begin{align}
\Omega _{\pm }\left( t\right) & =-\frac{{\omega_{B}}\left( t\right) }{2}\pm
\Omega \left( t\right) \nonumber\\ & =-\frac{{\omega_{B}}}{2}\pm \sqrt{\frac{{\omega_{B}}^{2}%
}{4}+{\omega_{E}}^{2}}\text{.}  \label{10b}
\end{align}
Since $\Omega _{-}\sim -\omega_{B}$ for ${\omega_{E}}\ll {\omega_{B}}$, the fast mode consists primarily of the particle cyclotron motion. On the other hand, since $\Omega _{+}\sim {\omega_{E}}^{2}/{\omega_{B}}=E/rB$ for ${\omega_{E}}\ll{\omega_{B}}$ the slow mode consists primarily of the drift motion. This ordering ${\omega_{E}}\ll {\omega_{B}}$ is relevant to most experiments because plasma instabilities arise when ${\omega_{E}}\sim{\omega_{B}}/2$ due to the large free energy content in the drift motion~\cite{Gueroult2017}. Going back from $Z\left( t\right) $ to the particle
position, one finally gets
\begin{widetext}
\begin{equation}
x+jy =\frac{\sqrt{\left| \Omega +\frac{3\Omega ^{\prime 2}}{8\Omega ^{3}}-%
\frac{\Omega ^{\prime \prime }}{4\Omega ^{2}}\right| _{0}}}{\sqrt{\left|
\Omega +\frac{3\Omega ^{\prime 2}}{8\Omega ^{3}}-\frac{\Omega ^{\prime
\prime }}{4\Omega ^{2}}\right| _{t}}}  
 \left[\rho \exp j\int_{0}^{t}\left( \Omega _{-}-\frac{3\Omega ^{\prime 2}%
}{8\Omega ^{3}}+\frac{\Omega ^{\prime \prime }}{4\Omega ^{2}}\right) du 
+R\exp j\int_{0}^{t}\left( \Omega _{+}+\frac{3\Omega ^{\prime 2}}{8\Omega
^{3}}-\frac{\Omega ^{\prime \prime }}{4\Omega ^{2}}\right) du\right]\text{,} 
\label{11} 
\end{equation}
\end{widetext}
with $\left( x=\rho +R,y=0\right) $ the initial position of the particle. Similarly to the static Brillouin rotation, the particle dynamics is now a superposition of the generalized fast ($\Omega _{-}$) and slow ($\Omega _{+}$) modes. 

Looking at Eq.~(\ref{11}), the classical Brillouin slow and fast rotations obtained for static $\mathbf{E}$ and $\mathbf{B}$ fields
are modified through two adiabatic effects. First, the amplitudes $%
\rho $ and $R$ are weakly modulated by the denominator $\sqrt{\Omega \left(
t\right) }$ $\sqrt{\left( 1+O\left[ \varepsilon ^{2}\right] \right) }$. As we will show next, 
this can be interpreted through the two classical adiabatic
invariants, the magnetic moment and the drift magnetic flux. Second, and of primary interest here, both the
fast rotation $\Omega _{-}<0$ and slow rotation $\Omega _{+}>0$ are sped up by an adiabatic factor $\left(
1+O\left[ \varepsilon ^{2}\right] \right)$. We will show in Sec.~\ref{Sec:SecIII} that the speed up of both the
fast cyclotron motion and the slow drift motion leads to the accumulation of
an adiabatic geometric phase. 

Before getting to the analysis of geometric phase effects, let us briefly review how the invariance of both the magnetic moment and the drift magnetic flux can be recovered from Eq.~(\ref{11}). 
To see this let us restrict ourselves to second order in $\varepsilon$ in the asymptotic expansion and assume that ${\omega_{E}}\ll {\omega_{B}}$ such that the drift and cyclotron dynamics are clearly separated. Consider then the product of $B(t)$ with the square of Eq.~(\ref{11}). Averaging out the slow motion, one gets
\begin{equation}
\frac{d}{dt}\left[ B\left( t\right)\left| x\left( t\right) +jy\left( t\right) \right| _{\Omega _{-}}^{2}\right] = 0,
\end{equation}
which expresses the magnetic flux invariance associated with the \textit{adiabatic invariance of the magnetic moment}. Averaging out the fast motion, one gets this times
\begin{equation}
\frac{d}{dt}\left[ B\left( t\right) \left| x\left(t\right) +jy\left( t\right) \right| _{\Omega _{+}}^{2}\right] = 0,
\end{equation}
which expresses the magnetic flux invariance associated with the \textit{adiabatic invariance of the flux encircled by the drift motion}: the ''\textit{third adiabatic invariant}'' of classical Alfv{\'e}n theory.

\section{Geometric phase of adiabatic Brillouin rotations}
\label{Sec:SecIII}

Let us now analyze the effect of a cyclic
variation of the control parameters of this Brillouin flow. Since the two control
parameters ${\omega_{B}}\left( t\right) $ and ${\omega_{E}}\left( t\right)$
are combined in a single physical variable $\Omega \left( t\right) $, one must turn to the control parameter phase space 
$\left( \Omega ,\Omega ^{\prime }\right) $ to look for an anholonomy effect. From Eq.~(\ref{11}), and recalling that $\Omega ^{\prime }dt=d\Omega $
and $\Omega ^{^{\prime \prime }}dt=d\Omega ^{\prime }$, the rotating phase
accumulated by a particle when the electric and
magnetic fields are varying between time $t$ and time $t+T$ writes
\begin{equation}
\int_{t}^{t+T}\Omega _{\pm }\left( u\right) du\pm \int_{\Omega \left(
t\right) }^{\Omega \left( t+T\right) }\frac{3\Omega ^{\prime }}{8\Omega ^{3}}%
d\Omega \mp \int_{\Omega ^{\prime }\left( t\right) }^{\Omega ^{\prime
}\left( t+T\right) }\frac{1}{4\Omega ^{2}}d\Omega ^{\prime }\text{.}
\label{12}
\end{equation}

\begin{figure}
\begin{center}
\includegraphics[width=8cm]{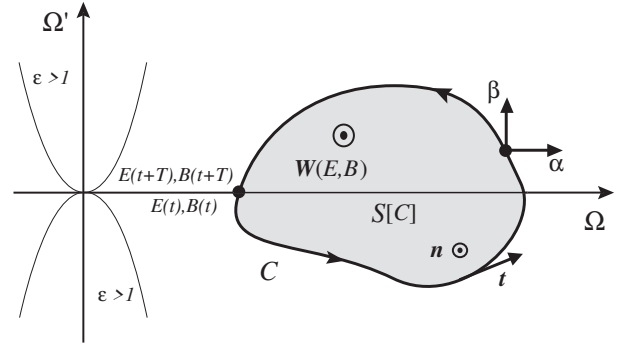}
\caption{The adiabatic domain ($\varepsilon <1)$ of the control parameter
phase space $\left( \Omega ,\Omega ^{\prime }\right) $. $C$ is a closed
curved travelled as a result of the cyclic adiabatic evolution of $E\left( t\right)$
and $B\left( t\right) $. $\mathbf{W}$  is the gauge field whose flux
through $S\left[ C\right] $ gives the accumulated geometric phase
associated with $C$.}
\label{Fig:Fig2}
\end{center}
\end{figure}

Consider the space $\left( \Omega ,\Omega ^{\prime }\right) $ as an
Euclidean vectorial space as illustrated in Fig.~\ref{Fig:Fig2}. Since the dot product of
\begin{equation}
\mathbf{\alpha }_{\pm }= \left( \pm 3\Omega ^{\prime }/8\Omega ^{3},0\right) 
\end{equation}
with $\left( d\Omega ,d\Omega ^{\prime }\right) $ is equal to $\pm
3\Omega ^{\prime }d\Omega /8\Omega ^{3}$, the second phase integral
in Eq.~(\ref{12}) can be viewed as a line integral of the vector $\mathbf{\alpha 
}_{\pm }$ along the parametric arc 
\begin{equation}
\left[ \Omega \left( u\right) ,\Omega ^{\prime
}\left( u\right) \right] _{t\leq u\leq t+T}
\end{equation}
drawn in the $\left( \Omega
,\Omega ^{\prime }\right) $ space. Similarly, since the dot product of 
\begin{equation}
\mathbf{\beta }_{\pm }=\left( 0,\mp 1/4\Omega ^{2}\right)
\end{equation}
with $\left( d\Omega ,d\Omega ^{\prime }\right) $ is equal to $\mp
d\Omega ^{\prime }/4\Omega ^{2}$, the third phase
integral in Eq.~(\ref{12}) can be viewed as a line integral of the vector $\mathbf{\beta }_{\pm }$ along the parametric arc $\left[ \Omega \left( u\right) ,\Omega ^{\prime }\left(
u\right) \right] _{t\leq u\leq t+T}$.

This geometric interpretation is particularly useful when, as considered here, the parametric
arc [$\Omega \left( u\right) $, $\Omega ^{\prime }\left( u\right) $] is a
closed curve $C$ corresponding to the cyclic adiabatic evolution of $B\left(
t\right) $ and $E\left( t\right) $ over a time $T$ such that
\begin{subequations}
\begin{equation}
\Omega\left( t+T\right)  = \Omega \left( t\right),
\end{equation}
\begin{equation}
\Omega ^{\prime }\left(t+T\right) = \Omega ^{\prime }\left( t\right)
\end{equation}
and 
\begin{equation}
\Omega ^{\prime
\prime }\left( t+T\right)  = \Omega ^{\prime \prime }\left( t\right).
\end{equation}
\end{subequations}
Under these conditions, Eq.~(\ref{11}) shows that the amplitudes $%
\rho $ $\sqrt{\Omega \left( 1+O\left[ \varepsilon ^{2}\right] \right) }$ and 
$R\sqrt{\Omega \left( 1+O\left[ \varepsilon ^{2}\right] \right) }$ are
unmodified by the cyclic adiabatic evolution. On the other hand, Eq.~(\ref{11})  shows that the angles of the slow and fast rotation increase by a quantity
\begin{equation}
\phi _{G}=\oint_{C}\left( \mathbf{\alpha }_{\pm }+\mathbf{\beta }_{\pm
}\right) \cdot \mathbf{t}ds=\pm \iint_{S\left[ C\right] }\frac{d\Omega
d\Omega ^{\prime }}{8\Omega ^{3}}  \label{13c}
\end{equation}
in addition to the expected classical dynamical phase 
\begin{equation}
\phi _{D} = \int_{t}^{t+T}\Omega _{\pm }du.
\end{equation}
Here 
\begin{equation}
\mathbf{t}=\left( d\Omega ,d\Omega^{\prime }\right) /\sqrt{d\Omega ^{2}+d\Omega ^{\prime 2}}
\end{equation} 
is the unit vector tangent to $C$, $ds$ = $\sqrt{d\Omega ^{2}+d\Omega ^{\prime 2}}$ is the
length element along $C$ and Stoke's relation was used in Eq.~(\ref{13c}%
) to transform the circulation along $C$ into a flux across $S\left[
C\right] $ the surface interior to $C$ in the Euclidean plane $\left( \Omega
,\Omega ^{\prime }\right) $ illustrated in Fig.~\ref{Fig:Fig2}. We note in passing here that $\left( \Omega ,\Omega ^{\prime }\right) $ can be viewed as the
phase space of the control parameter $\Omega $, and $C$ as a phase portrait
of the dynamics $\Omega \left( t\right) $.

With these results in hand, we conclude that the cyclic adiabatic evolution of the
control parameters $B\left( t\right)$ and $E\left( t\right)$ of a Brillouin rotation along a closed curve $C$ does not alter the drift radius $R$ of the slow rotation or the Larmor radius $\rho$ of the fast rotation. This result was expected on the basis of energy
conservation. Indeed, $R$ is associated with the potential energy of
the $\mathbf{E}$ field and $\rho $ is determined by the kinetic energy
of the cyclotron motion. The transformation along $C$ is cyclic and
adiabatic without resonances, thus this transformation is reversible and
there is no energy exchange and no energy dissipation after a full turn
along $C$. On the other hand, we find that the cyclic adiabatic evolution of the
control parameters introduces a
geometric phase to both the guiding center drift rotation and the cyclotron rotation. This
additional geometric phase is given by the flux of the vector
\begin{equation}
\mathbf{W}\left( E,B\right) =\left( {\omega_{B}}^{2}+4{\omega_{E}}^{2}\right)
^{-\frac{3}{2}}\mathbf{n}\text{,}  \label{14}
\end{equation}
through the surface $S$ inside $C$,  with $\mathbf{n}$ the unit vector normal to the Euclidean plane $\left(\Omega ,\Omega ^{\prime }\right) $ as shown in Fig.~\ref{Fig:Fig2}. Note also that $\mathbf{%
W}\left( E,B\right) $ is only a function of the value of the fields $\mathbf{E}$ and $\mathbf{B}$ and not of their derivatives. With this definition, the geometric
phase accumulated over the cyclic adiabatic evolution of the
control parameters $B\left( t\right)$ and $E\left( t\right)$ writes
\begin{widetext}
\begin{multline}
\left. \frac{x\left( T\right) +jy\left( T\right) }{x(0)+jy(0)}\right| _{%
\text{\textit{along} }C} =\frac{\rho }{R+\rho }\exp \left(
j\int_{0}^{T}\Omega _{-}dt\right)  \exp \left( -j\iint_{S\left[ C\right] }\mathbf{W}\cdot \mathbf{n}%
dS\right) \\+\frac{R}{R+\rho } 
\cdot \exp \left( j\int_{0}^{T}\Omega _{+}dt\right) \exp \left(
j\iint_{S\left[ C\right] }\mathbf{W}\cdot \mathbf{n}dS\right) \text{,}\label{15}
\end{multline}
\end{widetext}
where $dS=d\Omega d\Omega ^{\prime }$ is the surface element in $\Omega $
phase space. The dynamical phase $\phi _{D}=\int_{0}^{T}\Omega _{\pm }dt$ is
proportional to the duration of the cyclic evolution, while the geometric phase 
$\phi _{G}=\pm \iint_{S\left[ C\right] }\mathbf{W}\cdot \mathbf{n}dS$ is
proportional to the surface of the closed cycle in control parameter phase space.

\section{Interpretation through guiding center drift theory}
\label{Sec:SecIV}

Since the identified geometric phase $\phi _{G}$ is associated with higher order adiabatic effects,
it stands to reason that this result should be recovered through a careful analysis of the
higher order terms of classical drift theory. 

\subsection{Second order azimuthal drift}

To confirm this conjecture, let us consider for simplicity the special case where the capacitive radial electric field is null (${\omega_{E}}=0$). We thus analyze in this section the effect of an adiabatic cyclic evolution of the magnetic field over a duration $T$, with 
${\omega_{B}}\left( 0\right) ={\omega_{B}}\left( T\right) $ and $\omega
_{B}^{\prime }\left( 0\right) ={\omega_{B}}^{\prime }\left( T\right)$. 
The lowest order term of the electric drift asymptotic expansion is the $\mathbf{E}_{i}$ cross $\mathbf{B}$ drift. The next order term is the first inertial term associated with the unsteady $\mathbf{E}_{i}\left( t\right) $ cross $\mathbf{B}\left( t\right) 
$ velocity. When $\mathbf{B}$ is static, this second order term is the classical polarization drift. Here however, both $\mathbf{E}_{i}$ and $\mathbf{B}$ are time dependent, and we thus have to consider the full expression
\begin{equation}
\mathbf{v}=\frac{\mathbf{E}_{i}\times \mathbf{B}}{B^{2}}-\frac{m}{q}\frac{d}{%
dt}\left( \frac{\mathbf{E}_{i}\times \mathbf{B}}{B^{2}}\right) \times \frac{%
\mathbf{B}}{B^{2}}\text{.}  \label{17}
\end{equation}

Consider the polar basis $\left( \mathbf{u}_{r},\mathbf{u}_{\theta
}\right)$ and the associated polar coordinates $\left( r,\theta \right)$. The electric and magnetic fields given by Eqs.~(\ref{1}) and (\ref{3}) and their projections on the polar basis leads to the relation
\begin{equation}
\frac{\mathbf{E}_{i}\times \mathbf{B}}{B^{2}}=-\frac{{{\omega_{B}}}^{\prime }%
\text{ }}{2{\omega_{B}}}r\mathbf{u}_{r}  \label{18}
\end{equation}
for the $\mathbf{E}_{i}$ cross $\mathbf{B}$ drift, and to the relation 
\begin{multline}
-\frac{m}{q}\frac{d}{dt}\left( \frac{\mathbf{E}_{i}\times \mathbf{B}}{B^{2}%
}\right) \times \frac{\mathbf{B}}{B^{2}} 
=\frac{{\omega_{B}}^{\prime }\text{ }}{2{\omega_{B}}^{2}}r\frac{%
d\theta }{dt}\mathbf{u}_{r}\\+\left( -\frac{{\omega_{B}}^{^{\prime \prime }}\text{ }}{2{\omega_{B}}^{2}}r+%
\frac{{\omega_{B}}^{\prime 2}\text{ }}{2{\omega_{B}}^{3}}r-\frac{{\omega_{B}}^{\prime }\text{ }}{2{\omega_{B}}^{2}}\frac{dr}{dt}\right) \mathbf{u}%
_{\theta }  \label{19}
\end{multline}
for the polarization drift. Writing
\begin{equation}
\mathbf{v}=\frac{dr}{dt}\mathbf{u}_{r}+r\frac{d\theta }{dt}\mathbf{u}%
_{\theta }  \label{16}
\end{equation}
the polar representation of the guiding center drift velocity of a particle of mass $m$ \ and charge $q$ and identifying the radial and azimuthal components of Eq.~(\ref{19}) and Eq.~(\ref{16}) then yields 
\begin{subequations}
\begin{align}
\frac{dr}{dt} =&-\frac{{\omega_{B}}^{\prime }\text{ }}{2{\omega_{B}}}%
r+O\left[ \varepsilon ^{3}\right] \text{,}  \label{20a} \\
\frac{d\theta }{dt} =&\frac{3{\omega_{B}}^{\prime 2}\text{ }}{4{\omega_{B}}^{3}}-\frac{{\omega_{B}}^{^{\prime \prime }}\text{ }}{2{\omega_{B}}^{2}}%
+O\left[ \varepsilon ^{4}\right] \text{.}  \label{20b}
\end{align}
\end{subequations}

Eq.~(\ref{20a}) is simply the invariance of the magnetic flux $%
{\omega_{B}}r^{2}$, also known as the \textit{third adiabatic invariant} of classical
Alfven' adiabatic theory. The guiding center radial position $r$ is not
affected by the cyclic adiabatic evolution $B\left( 0\right) =B\left(
T\right) $, and the solution of Eq.~(\ref{20a}) is 
\begin{equation}
\sqrt{{\omega_{B}}\left( T\right) }r\left( T\right) =\sqrt{{\omega_{B}}\left(
0\right) }r\left( 0\right) \text{.}  \label{21}
\end{equation}
On the other hand, the polar angle of the guiding center is affected by the cyclic
adiabatic evolution $B\left( 0\right) =B\left( T\right) $. The polar angle variation is obtained by integrating Eq.~(\ref{20b}), 
\begin{equation}
\theta \left( T\right) =\theta \left( 0\right) +\int_{0}^{T}\left( 
\frac{3{\omega_{B}}^{\prime 2}\text{ }}{4{\omega_{B}}^{3}}-\frac{{\omega_{B}}^{^{\prime \prime }}\text{ }}{2{\omega_{B}}^{2}}\right) du\text{.}
\label{22b}
\end{equation}
As anticipated, this is exactly the result predicted in Eq.~(\ref{11}) once noted
that $\Omega \left( t\right) ={\omega_{B}}\left( t\right) /2$ should be used instead of the general definition Eq.~(\ref{7}) since we considered here ${\omega_{E}}=0$.

\subsection{Origin of phase accumulation through an ideal cycle}

\begin{figure}
\begin{center}
\includegraphics[width=8cm]{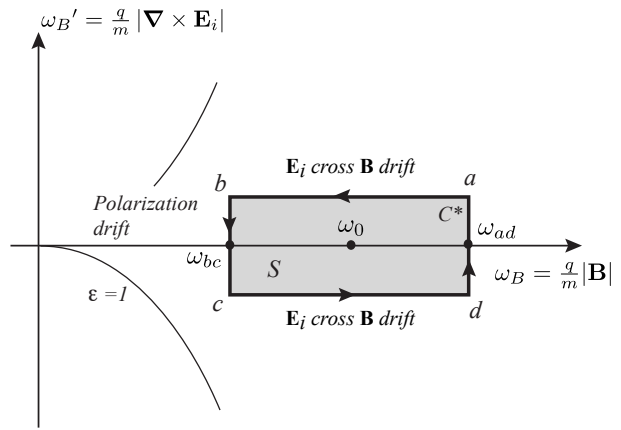}
\caption{Simple rectangular adiabatic cycle $\left( a,b,c,d\right) $ in $%
\left( {\omega_{B}},{\omega_{B}}^{\prime }\right) $ space.}
\label{Fig:Fig3}
\end{center}
\end{figure}

\begin{figure}
\begin{center}
\includegraphics[width=7cm]{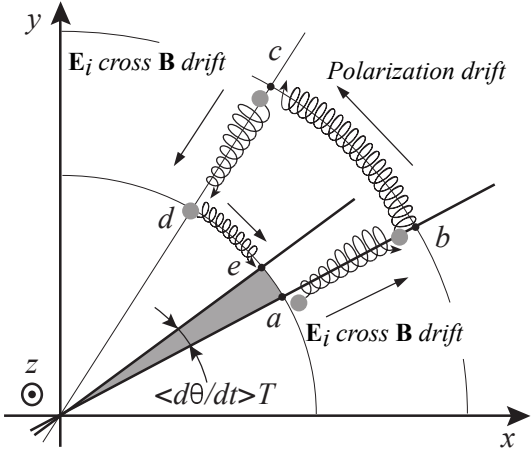}
\caption{Real space $\left( x,y\right) $ effects of the closed circuit $C^{*}$
in $\left( {\omega_{B}},{\omega_{B}}^{\prime }\right) $ space. The real space
circuit is not closed and displays an angular shift after each completion of the $C^{*}$ cycle.}
\label{Fig:Fig4}
\end{center}
\end{figure}

This analysis through the particle drift motion can also offer insights into the origin of phase accumulation. To see this, consider the simple cyclic adiabatic evolution $C^{*}$ depicted in Fig.~\ref{Fig:Fig3}. It corresponds to a rectangle of surface $S$, centered at (${\omega_{B}}={\omega_0}$, ${\omega_{B}}^{\prime }=0$),  in the control parameter phase space 
\begin{equation}
({\omega_{B}} = \frac{q}{m}\left| \mathbf{B}\right|, {\omega_{B}}^{\prime } = \frac{q}{m}\left| \mathbf{\nabla }%
\times \mathbf{E}_{i}\right|),
\end{equation}
where it is further assumed that ${\omega_{bc}}<{\omega_0}<{\omega_{ad}}$. We note that this cycle does not represent a physical solution since, as we will show, $B$ is in this case non-differentiable. Yet, this configuration makes it possible to capture in a simple manner the basic features of phase accumulation. We thus focus here on this ideal cycle, and leave the study of a more practical cycle $C_{h}$ to Sec.~\ref{Sec:SecV}. The four corners of the rectangle $C^{*}$ are labelled $a$, $b$, $c$ and $d$. The transit time from $a$ to $b$ is assumed to be equal to the transit time from $c$ to $d$, while the transit time from $b$ to $c$ is assumed to be equal to that from $d$ to $a$. The duration of a complete turn around ${\omega_0}$ along $C^{*}$ is $T$ such that ${\omega_0}T\gg 1$. For a continuous evolution along $C^{*}$ around ${\omega_0}$, Eq.~(\ref{13c}) shows that the average angular velocity of a particle is
\begin{equation}
\left\langle \frac{d\theta }{dt}\right\rangle =\frac{S}{8}\frac{{\omega_0}}{%
\left( {\omega_{bc}}{\omega_{ad}}\right) ^{2}T}\text{.}  \label{23}
\end{equation}
To understand the phenomena leading to this result, let us consider sequentially the particle dynamics associated with each of the four sides of this closed curve. The corresponding particle trajectory in real space is shown in Fig.~\ref{Fig:Fig4}.

Starting at time $t=0$, the magnetic field $\mathbf{B}$ decreases linearly with time from $a$ to $b$. As a result, the inductive field $\mathbf{E}_{i}$ given in Eq.~(\ref
{3}) is independent of time and the $\mathbf{E}_{i}$ cross $\mathbf{B}$ drift
associated with this azimuthal inductive electric field drives a radial
motion. Concurrently to this outward radial drift motion at constant polar
angle, the Larmor radius increases in order to fulfil the adiabatic
invariance of the magnetic moment. From $b$ to $c$, the magnetic field $\mathbf{B}$ is constant $\left(|\mathbf{B}| =  m\omega
_{bc}/q\right)$. On the other hand, the inductive electric field $\mathbf{E}_{i}$ exhibits a very
large time variation which becomes the source of an azimuthal polarization as illustrated in Fig.~\ref{Fig:Fig4}. Due to the discontinuity of the time derivative of $B\left( t\right)$ in our ideal cycle, the inductive electric field in $b$ and $c$ should be understood here as the left and right derivative of $B\left( t\right)$, respectively. Then, from $c$ to $d$, the magnetic field $\mathbf{B}$ increases linearly at constant
inductive electric field $\mathbf{E}_{i}$. This leads to an inward radial $\mathbf{E}_{i}$ cross $\mathbf{B}$
drift exactly opposed to the $a$ to $b$ transit. It is simply the time
reversal of $a$ to $b$ governed by Eq.~(\ref{21}). Finally, the segment from $d$ to $a$ completes the circuit in the $\left(\Omega ,\Omega ^{\prime }\right) $ phase space depicted in Fig.~\ref{Fig:Fig3}. In contrast,
Fig.~\ref{Fig:Fig4} shows that the real space $\left( x,y\right)$ circuit remains open. The position at $t=T$ after the complete cyclic evolution of $\mathbf{B}$ is not $a$
but $e\neq a$. This $e-a$ mismatch is due to the fact that the polarization drift
along this last segment of the circuit corresponds to the same $\left| dE_{i}/dt\right| $ as the $b$ to $c$ segment, but with a stronger magnetic field $\left( B=m{\omega_{ad}}/q>m{\omega_{bc}}/q\right)$. Indeed, Eq.~(\ref{21}) indicates that
\begin{equation}
\frac{r_a}{r_b}=\frac{r_d}{r_c}\sim \sqrt{\frac{B_b}{B_a}}=\sqrt{\frac{B_c}{B_d}}
\end{equation}
so that the arc $\left( d,a\right) $ is shorter than the arc $\left( b,c\right)$ by a factor equal to the square root of the magnetic field ratio. Meanwhile, the polarization drift is proportional to $B^{-2}$. Putting these two results together, the polarization drift velocity is too small by a factor proportional to $\left( B_{c}/B_{d}\right) ^{-3/2}$ for particles to reach $a$.

This sequence of alternate $\mathbf{E}_{i}$ cross $\mathbf{B}$ drifts and polarization
drifts provides a clear physical picture of the origin of the geometric
phase in the simple case ${\omega_{E}}=0$.

\section{Adiabatic cycle as a means to drive plasma rotation}
\label{Sec:SecV}

As underlined above, the $C^{*}$ cycle studied in Sec.~\ref{Sec:SecIV} is
not physical in that it corresponds to a non-differentiable magnetic field. To address this issue, we consider first in this section an elliptical cycle $C_{h}$. This cycle is the simplest possible periodic variation, both from theoretical and practical standpoints. 

\subsection{Extension to a physical cycle}

\begin{figure}
\begin{center}
\includegraphics[width=8cm]{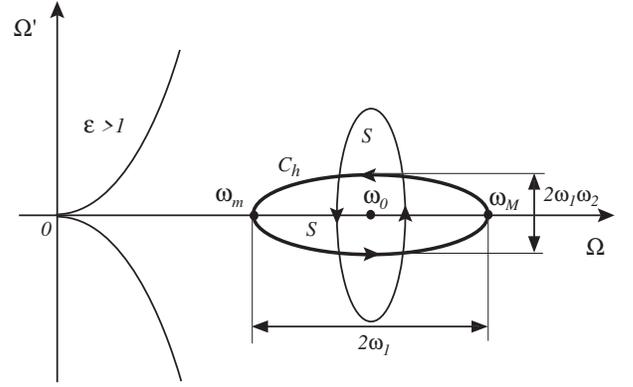}
\caption{Closed elliptical curves $C_{h}$ in $\left( \Omega ,\Omega ^{\prime
}\right) $ space associated with an harmonic modulation of $\Omega $.}
\label{Fig:Fig5}
\end{center}
\end{figure}

Consider an harmonic swing of the control parameter
\begin{equation}
\Omega \left( t\right) ={\omega_0}+{\omega_1}\cos \left( \omega_{2}t\right) \text{,}  \label{24}
\end{equation}
with angular frequency $\omega_2$ around a central value ${\omega_0}$, with $\omega _{1}=({\omega_M}-{\omega_m})/2$. With these definitions, the adiabaticity condition writes
\begin{equation}
\varepsilon =\frac{\left| {\omega_1}{\omega_2}\right|}{{\omega_0}^{2}}\ll 1
\end{equation}
and the closed curve $C_{h}$ associated with this
harmonic variation is the ellipse 
\begin{equation}
\left( \frac{\Omega -{\omega_0}}{{\omega_1}}\right) ^{2}+\left( \frac{%
\Omega ^{\prime }}{{\omega_1}{\omega_2}}\right) ^{2}=1  \label{25}
\end{equation}
depicted in Fig.~\ref{Fig:Fig5}. The surface $S$ of $C_{h}$ is given by the relation $S=\pi {\omega_{1}}^{2}{\omega_2}$.

In one complete cycle along the ellipse $C_{h}$, Eq.~(\ref{13c}) predicts that the slow and fast Brillouin drift and cyclotron modes accumulate a geometric phase  
\begin{equation}
\iint_{S\left[ C_{h}\right] }\mathbf{W}\cdot \mathbf{n}dS=\iint_{S\left[
C_{h}\right] }\frac{d\Omega d\Omega ^{\prime }}{8\Omega ^{3}}=\frac{S}{%
8\left( \omega _{m}\omega _{M}\right) ^{\frac{3}{2}}}\text{.}  \label{26}
\end{equation}
The steady state slow oscillation Eq.~(\ref{24}) is responsible for an
averaged angular rotation $\left\langle d\theta /dt\right\rangle $ of both
the cyclotron phase and the guiding center drift phase. For the latter,
the drift angular frequency is a function of the surface $S$, and 
\begin{equation}
\left\langle \frac{d\theta }{dt}\right\rangle =\frac{S}{16\pi }\frac{\omega
_{2}}{\left( \omega _{m}\omega _{M}\right) ^{3/2}}\approx \frac{{\omega_{1}}^{2}{\omega_2}^{2}}{16{\omega_0}^{3}}\text{.}  \label{27b}
\end{equation}
Although Eq.~(\ref{27b}) appears to suggest a different frequency scaling when compared to Eq.~(\ref{23}), this apparent inconsistency is resolved when considering the geometry of the rectangle $C^{*}$ and the ellipse $C_{h}$. Indeed, from Fig.~\ref{Fig:Fig4} and Fig.~\ref{Fig:Fig5}, one identifies $\left.
\omega _{m}\omega _{M}\right| _{C_{h}}\sim \left. {\omega_{bc}}\omega
_{ad}\right| _{C^{*}}$ and $\left. {\omega_2}\sqrt{\omega _{m}\omega _{M}}%
\right| _{C_{h}}\sim \left. {\omega_0}/T\right| _{C^{*}}$.

\subsection{Rotation drive and equivalent electric field}

Practically, the elliptical cycle studied in the first part of this section can be realized through a weak modulation of the axial magnetic field
\begin{equation}
\mathbf{B}\left( t\right) =\left[ B+b\cos (\omega t)\right] \mathbf{u}_{z}%
\text{.}  \label{28b}
\end{equation}
To satisfy to the the adiabatic condition, we further require that $\omega \ll {\omega_{B}}=qB/m$ and $b/B\ll 1$. Under these conditions, the control parameter writes 
\begin{equation}
\Omega =\frac{{\omega_{B}}}{2}\left[1+\frac{b}{B}\cos (\omega t)\right],
\end{equation}
and identification with Eq.~(\ref{24}) yields $\omega_0 = {\omega_{B}}/2$, $\omega_1 = {\omega_{B}b}/(2B)$ and $\omega_2 = \omega$. From Eq.~(\ref{27b}), the adiabatic modulation of $\mathbf{B}$ thus leads to a guiding center rotation with drift angular frequency 
\begin{equation}
\varpi = \frac{1}{8}\left(\frac{b}{B}\right)^2\frac{\omega^2}{\omega_{B}}.
\label{Eq:omega_B_mod}
\end{equation}
The weak and adiabatic modulation of the axial magnetic field therefore drives a rigid body rotation. Quantitatively, since the adiabatic condition requires both $\omega \ll {\omega_{B}}$ and $b/B\ll 1$, Eq.~(\ref{Eq:omega_B_mod}) shows that $\varpi\ll\omega$. This rotation is therefore slow.

Looking for analogies with the classical static Brillouin flow, one notices that the azimuthal drift angular frequency Eq.~(\ref{Eq:omega_B_mod}) is identical to the angular frequency $\varpi = |\mathbf{E^{\star}}\times\mathbf{B}|/r$ which would be found in the presence of a capacitive radial electric field
\begin{equation}
\frac{\mathbf{E}^{*}}{B}=\left( \frac{b}{B}\right) ^{2}\left( \frac{\omega }{%
{\omega_{B}}}\right) \frac{\omega }{8}\left( x\text{ }\mathbf{u}_{x}+y\text{ }%
\mathbf{u}_{y}\right) \text{.}  \label{29}
\end{equation}
However, the small plasma angular frequency predicted for this scheme suggests that this configuration does not enjoy the same upside potential for mass separation applications as that of classical Brillouin rotating configurations in which centrifugal effects may be heightened~\cite{Gueroult2018}. Yet, a closer look reveals possible new opportunities through the mass dependence of $\mathbf{E}^{*}$, which will be studied in detail in a forthcoming paper.

\section{Summary and discussion}
\label{Sec:SecVI}

We have extended through an adiabatic expansion the classical Brillouin rotation modes to the regime where the driving crossed electric and magnetic fields are no longer static but slowly evolve in time. By considering the phase space of the control parameter $\Omega \left(E,B\right) $ instead of the state space of this control parameter, a geometric phase is shown to arise from the cyclic adiabatic variation of $E$ and $B$, and this phase effect is then traced back to a higher order guiding center drift. The main new results of this study are given by Eqs.~(\ref{15}), (\ref{26}) and (\ref{29}) and the physical analysis of Sec.~\ref{Sec:SecIV}.

This finding of a geometrical phase effect in Brillouin rotations was then used to show that a weak adiabatic modulation of the axial magnetic leads to a slow rotation of the guiding center around the magnetic axis, and that this configuration is equivalent for guiding center dynamics to a crossed field configuration with a DC capacitive radial electric field $\mathbf{E^{\star}}$ given in Eq.~(\ref{29}). This conceptual solution to drive rotation appears promising and advantageous given the challenges typically associated with crossed field driven rotation~\cite{Gueroult2019}. 

Although the mass dependence of $\mathbf{E^{\star}}$ offers promise for mass separation and deserves further attention, some possible challenges can already be brought forward. One will for instance have to  account for skin effects when considering magnetic field penetration in dense plasma columns of large radii. Indeed, one expects the modulating field $b$ to no longer be uniform, but instead defined by Bessel functions for plasma radii larger than the skin depth. This question, as well as the analysis of designs which could remediate to some of these limitations, is left for a future study dedicated to the practical applications of this geometric phase $\phi _{G}$ and equivalent electric field $\mathbf{E}^{*}$.

\section*{References}
%

\end{document}